\documentclass{article}

\usepackage{PRIMEarxiv}
\usepackage{amssymb}
\usepackage[T1]{fontenc}
\usepackage{graphicx}
\usepackage{color}
\usepackage{color, soul}
\usepackage{microtype}
\soulregister\cite7
\soulregister\ref7
\soulregister\pageref7
\usepackage{multirow}
\usepackage{comment}
\usepackage{amssymb}
\usepackage{pifont}
\usepackage{amsmath} 
\usepackage[linesnumbered,ruled,vlined]{algorithm2e}
\usepackage{caption}
\usepackage{subcaption}
\usepackage{dsfont}
\usepackage{url}
\usepackage{subcaption}

\usepackage{booktabs}
\title{Towards Certified Malware Detection: Provable Guarantees Against Evasion Attacks}

\author{
Nandakrishna Giri\\
Department of Computer Applications, \\
Cochin University of Science \\
and Technology, India \\
\texttt{nandakrishnagiri@pg.cusat.ac.in} \\
\And
Asmitha K. A.\\
Department of Computer Applications, \\
Cochin University of Science \\
and Technology, India \\
\texttt{asmitha@pg.cusat.ac.in} \\
\And
Serena Nicolazzo \\
Department of Electrical, Computer\\
and Biomedical Engineering, University of Pavia, \\
Via Adolfo Ferrata 5, 27100, Pavia, Italy\\
\texttt{serena.nicolazzo@unipv.it} \\
\And
Antonino Nocera\\
Department of Electrical, Computer\\
and Biomedical Engineering, University of Pavia, \\
Via Adolfo Ferrata 5, 27100, Pavia, Italy\\
\texttt{antonino.nocera@unipv.it} \\
\And
Vinod P. \\
Department of Computer Applications, \\
Cochin University of Science \\
and Technology, India \\
\texttt{vinod.p@cusat.ac.in} \\
}

\begin{document}

\maketitle

\date{April 2026}

\begin{abstract}
Machine learning–based static malware detectors remain vulnerable to adversarial evasion techniques, such as metamorphic engine mutations. To address this vulnerability, we propose a certifiably robust malware detection framework based on randomized smoothing through feature ablation and targeted noise injection. During evaluation, our system analyzes an executable by generating multiple ablated variants, classifies them by using a smoothed classifier, and identifies the final label based on the majority vote. By analyzing the top-class voting distribution and the Wilson score interval, we derive a formal certificate that guarantees robustness within a specific radius against feature-space perturbations. We evaluate our approach by comparing the performance of the base classifier and the smoothed classifier on both clean executables and ablated variants generated using PyMetaEngine. Our results demonstrate that the proposed smoothed classifier successfully provides certifiable robustness against metamorphic evasion attacks without requiring modifications to the underlying machine learning architecture.
\end{abstract}


\section{Introduction}

Malware detection has become increasingly challenging due to the rapid evolution of evasion techniques and the growing complexity of modern executable files. Traditional machine learning approaches, while effective under standard conditions, often struggle to maintain robustness when faced with adversarially manipulated or structurally modified inputs.
Unlike conventional image or natural language processing tasks, malware detection operates in a fully adversarial environment. Malware authors continuously modify their binaries to evade detection. Typically, adversarial attacks involve carefully crafted modifications that preserve functionality while aiming to force the ML model to misclassify a malicious file as benign \cite{abaid2017quantifying}.

Classic machine learning models for static malware detection rely on structural features extracted from executables. For example, the EMBER dataset and its extraction pipeline provide an effective benchmark for parsing Portable Executable (PE) features for classification \cite{anderson2018ember}. The PE file format describes the predominant executable
format for Microsoft Windows operating systems, and includes executables, dynamically-linked libraries (DLLs), and
FON font files. However, attackers can exploit these models by using metamorphic engines, such as PyMetaEngine\footnote{\url{https://github.com/scmanjarrez/pymetangine}}, which reconstructs the binary assembly without altering the underlying malicious payload. These metamorphic transformations shift the resulting feature vector just enough to cross the model's decision boundary, making classifiers vulnerable to evasion. Given these vulnerabilities, there is a critical need for defenses that provide formal mathematical guarantees rather than relying solely on empirical robustness. One promising approach is randomized smoothing \cite{cohen2019certified}, which transforms an arbitrary base classifier into a robust smoothed classifier by aggregating predictions over noise-corrupted inputs. Subsequent research has extended smoothing to feature ablation \cite{levine2020randomized} to certify robustness against patch-based attacks. In ablation-based smoothing, a classifier is trained on heavily masked inputs, and test-time predictions are obtained via majority voting over multiple perturbed copies.

Building on these certifiable defenses, we propose a novel smoothed classification scheme specifically designed for the discrete feature space of malware executables. During training, features are clustered into discrete groups, and a fixed percentage of these groups is retained, while the rest are set to zero. Additionally, random noise is injected into a subset of the surviving groups. At inference time, the smoothed classifier generates multiple randomly ablated versions of the input and makes the final prediction based on the majority vote. By tracking the top-class voting distribution and applying the Wilson score interval, we derive a robustness certificate that guarantees the classification remains stable within a specific perturbation radius.
Hence, our novel contribution consists of adapting and extending randomized smoothing to the discrete feature space of malware detection by combining feature grouping, ablation, and targeted noise injection, and enabling certified robustness against realistic metamorphic evasion attacks.

To systematically evaluate the efficacy of our proposed defense, this research addresses the following core questions:
\begin{itemize}
\item RQ1: To what extent do metamorphic evasion techniques degrade the detection performance of standard feature-based malware classifiers?
\item RQ2: Does applying feature-based randomized smoothing, using ablation and noise injection, increase empirical robustness against metamorphic attacks compared to an unprotected baseline?
\item RQ3: How can the robustness of the smoothed classifier be mathematically certified against adversarial feature perturbations using the majority voting mechanism and the Wilson score interval?
\item RQ4: Does the proposed feature-space smoothing methodology generalize across different machine learning architectures, specifically deep learning models?
\end{itemize}

Building on these research questions, we present the following key contributions:

\begin{itemize}
\item We propose a randomized smoothing framework that combines feature grouping, ablation, and noise injection specifically for feature-based malware detection.
\item We provide a mathematical mechanism to calculate certifiable robustness guarantees for malware feature perturbations using majority voting and the Wilson score interval.
\item We empirically evaluate our certified defense against metamorphic evasion attacks generated via PyMetaEngine, demonstrating that the proposed smoothed classifier is more robust to adversarial modifications than a standard, unprotected base classifier.
\end{itemize}

The remainder of this paper is organized as follows. Section \ref{sec:related} reviews the most relevant work in this area. Section \ref{sec:method} presents the main steps of our proposed framework, while Section \ref{sec:result} reports its results. Finally, Section \ref{sec:conclusion} concludes the paper and outlines directions for future work.

\section{Related Work}
\label{sec:related}

Randomized smoothing has emerged as a prominent technique for achieving certified robustness in machine learning models. Cohen et al.~\cite{cohen2019certified} demonstrated that a smoothed classifier constructed via Gaussian noise injection can provide formal robustness guarantees under L2-bounded adversarial perturbations. However, this approach is primarily designed for continuous input domains and relies on additive Gaussian noise, which is not directly applicable to the discrete and structured feature space of malware detection. 

Recent work has further improved randomized smoothing by optimizing the training process to better balance the trade-off between accuracy and robustness. In particular, sample-wise training strategies have been proposed to selectively enhance robustness based on the sensitivity of individual inputs to Gaussian noise \cite{jeong2023confidence}. While this method improves certified robustness in continuous domains, it remains limited to image-based settings and does not directly address the challenges posed by discrete and structured malware features.
Alternative approaches to certified robustness have been proposed based on Differential Privacy (DP), for instance, the defense mechanism of PixelDP \cite{lecuyer2018certified}. While effective and scalable, such an approach is primarily designed for continuous input domains and do not directly address the discrete and structured nature of malware features.
While these methods demonstrate the efficacy of smoothed classifiers in image domains, their adaptation to malware detection remains largely unexplored.
Some recent work has extended certified robustness to more general architectures and training strategies. For example, Li et al.~\cite{li2019certified} introduced a framework connecting robustness to adversarial perturbations with additive noise and proposed a training strategy that improves certified bounds for general network architectures.

Recent proposals adapt randomized smoothing to the malware detection context \cite{gibert2023certified,gibert2024adversarial,saha2023drsm,gao2024certrob}. In particular, Gibert et al.~\cite{gibert2023certified,gibert2024adversarial} proposed chunk-based smoothing approaches for malware detection. While \cite{gibert2023certified} introduces a certifiable defense against patch and append attacks, providing formal robustness guarantees, the work \cite{gibert2024adversarial} focuses on empirical robustness against general adversarial malware manipulations with flexible chunk selection strategies. In contrast, our approach operates on higher-level feature groups with targeted ablation and noise injection, enabling certified robustness specifically against realistic metamorphic evasion attacks, bridging the gap between empirical and provable defenses in discrete feature spaces.
Always in the field of malware detection, the authors of \cite{saha2023drsm} propose DRSM, a certified malware detection framework based on de-randomized smoothing applied to MalConv. Our approach operates on higher-level feature groups with targeted ablation and noise injection, enabling certified robustness specifically against metamorphic evasion attacks across arbitrary ML classifiers. Whereas, Gao et al.~\cite{gao2024certrob} propose CertRob, a PDF malware detection framework with certified adversarial robustness using randomized smoothing. Their approach applies feature-constrained smoothing and Shapley value–based prioritization to defend against both feature-space and problem-space attacks. In contrast, our work focuses on PE executables, leveraging feature grouping, targeted ablation, and noise injection to provide certified robustness specifically against metamorphic evasion attacks in discrete feature spaces.

In contrast to prior work, we extend randomized smoothing to the discrete feature space of malware by combining feature grouping, controlled ablation, and targeted noise injection. We further introduce a certification mechanism based on majority voting and the Wilson score interval, enabling formal robustness guarantees within a bounded perturbation radius. Our approach specifically targets realistic metamorphic evasion scenarios, bridging the gap between empirical robustness and provable security in malware detection. A comparative summary of existing methods across image and malware domains is provided in Table~\ref{tab:comparison}.

\begin{table}[t]
\centering
\scriptsize
\caption{Comparison of robustness approaches in malware detection}
\label{tab:comparison}
\resizebox{\textwidth}{!}{
\begin{tabular}{l|llllp{1.6cm}p{2cm}}
\toprule
\textbf{Ref.} & \textbf{Year} &  \textbf{Domain} & \textbf{Input} & \textbf{Perturbation} & \textbf{Certified Robustness} & \textbf{Attack Model} \\
\midrule

Lecuyer et al.~\cite{lecuyer2018certified} & 2018 & Images & Continuous 
& Noise (DP-based)
& \checkmark & Norm-bounded adversarial \\

Cohen et al.~\cite{cohen2019certified} & 2019
& Images & Continuous 
& Gaussian noise  
& \checkmark & L2 adversarial \\

Li et al.~\cite{li2019certified} & 2019 & Images & Continuous & Additive noise & \checkmark & L1 adversarial\\

Gibert et al. \cite{gibert2023certified} & 2023 & Malware & Bytes & Random ablation 
& \checkmark & Patch/ Append Attacks) \\

Jeong et al.~\cite{jeong2023confidence} & 2023 & Images & Continuous & Gaussian noise & \checkmark & L2 adversarial \\

Saha et al. \cite{saha2023drsm} & 2023 & Malware & Bytes & Window ablation & \checkmark & Contiguous byte evasion \\

Gao et al.~\cite{gao2024certrob} & 2024 & Malware & Features & Randomized smoothing + feature constraints & \checkmark & Problem-space attacks \\

Gibert et al.~\cite{gibert2024adversarial} & 2024 & Malware & Bytes & Chunk-based smoothing & - & Adversarial malware evasion \\

\hline
Ours & 2026 & Malware 
& Features & Ablation + noise 
& \checkmark
& Metamorphic evasion \\

\bottomrule
\end{tabular}
}
\end{table}

\section{Methodology}
\label{sec:method}

In this section, we present our proposed methodology in detail, describing both {\em Training} and {\em Inference phases} of our approach.

\subsection{Training Phase}
\begin{figure}[htbp]
    \centering
    \includegraphics[width=\textwidth]{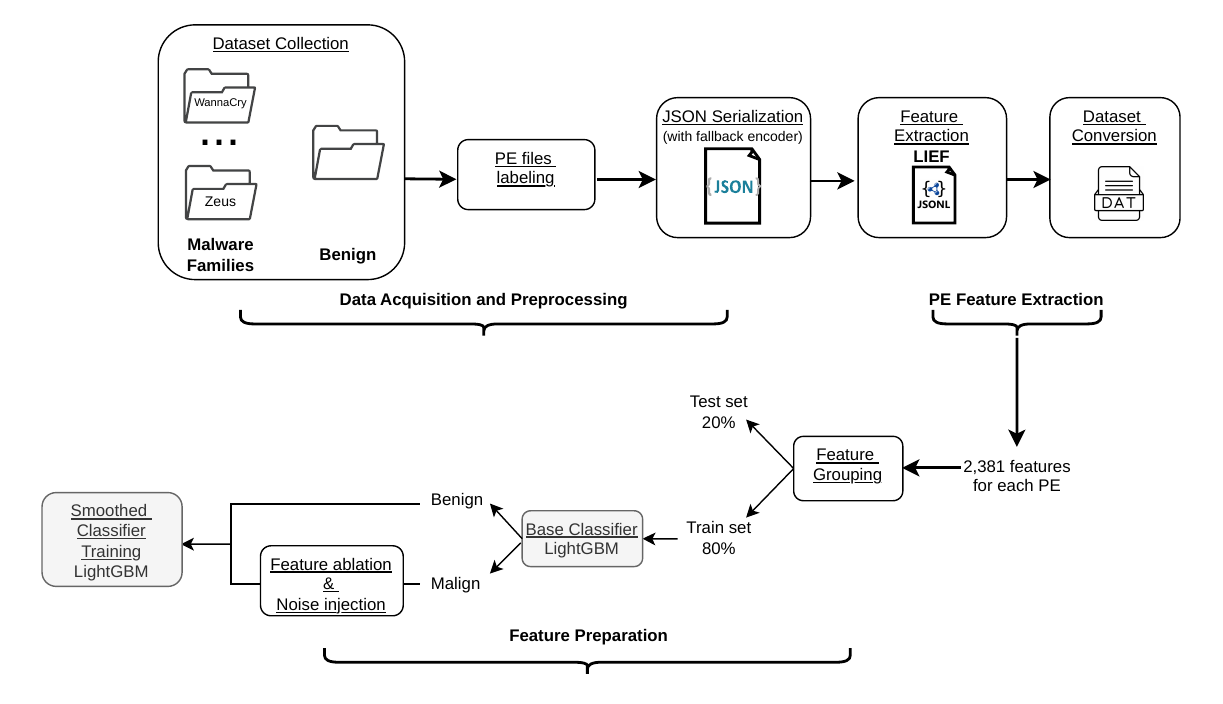}
    \caption{Complete Training Pipeline of our approach}
    \label{fig:architecture}
\end{figure}

As illustrated in Figure \ref{fig:architecture}, the {\em Training Phase} of our framework consists of the following key steps:

\begin{itemize}
    \item \textbf{Data Acquisition and Preprocessing}. We collect a dataset of Portable Executable (PE) files from multiple sources, labeling each sample as either malicious or benign. Preprocessing steps are applied to ensure data quality, including the removal or correction of corrupted or inconsistent files.

    \item \textbf{PE Feature Extraction}. For each PE file, we extract a set of $2,381$ structural features, capturing relevant characteristics of the executable for downstream classification.

    \item \textbf{Feature Preparation}. The extracted features are processed to fit the proposed smoothed classification framework. In particular, features are grouped into semantically meaningful subsets, followed by the application of feature ablation and targeted noise injection to support robustness through randomized smoothing.

   \item \textbf{Smoothed Classifier Training}. Finally, the prepared feature representations are used to train the smoothed classifier. The model is trained on multiple randomly perturbed versions of the input data, enabling it to learn robust decision boundaries.

\end{itemize}

In the following subsections, we provide a detailed description of each of these steps.

\subsubsection{Data Acquisition and Preprocessing.}

The first step of our framework consists of constructing a labeled dataset of Portable Executable (PE) files. Let $\mathcal{D}$ denote the complete dataset, defined as the union of two disjoint subsets:
\[
\mathcal{D} = \mathcal{D}_m \cup \mathcal{D}_b, \quad \mathcal{D}_m \cap \mathcal{D}_b = \emptyset,
\]
where: 

\begin{itemize}
    \item $\mathcal{D}_m$ represents the set of malicious samples from the Malhub dataset \cite{asmitha2024deep}, formally,
    $\mathcal{D}_m = \{ x_i \mid x_i \in \text{PE},\; y_i = 1,\; i = 1,\dots,N_m \}$, where $N_m = 26{,}452$ samples spanning multiple malware families (e.g., WannaCry, Zeus), collected from the VirusShare repository\footnote{\url{https://virusshare.com}}.

    \item $\mathcal{D}_b$ the set of benign samples is defined as $\mathcal{D}_b = \{ x_j \mid x_j \in \text{PE},\; y_j = 0,\; j = 1,\dots,N_b \}$ and it is a collection of various sources, such as the Windows operating system and GitHub repositories, including PE samples from  PE Malware Machine Learning Dataset\footnote{\url{https://practicalsecurityanalytics.com/pe-malware-machine-learning-dataset}} for the benign PE files.
\end{itemize}

For the purpose of model training, we established a consistent labeling scheme by assigning a binary label to each sample. After this, each executable $x \in \mathcal{D}$ is associated with a binary label function:
\[
y: \mathcal{D} \rightarrow \{0,1\}, \quad
y(x) =
\begin{cases}
1 & \text{if } x \in \mathcal{D}_m, \\
0 & \text{if } x \in \mathcal{D}_b.
\end{cases}
\]

Due to the continuous use of obfuscation techniques by malware authors, such as the manipulation of PE headers and the injection of extraneous data to evade analysis, feature extraction may encounter malformed or corrupted data.  To address this issue, we implemented a custom fallback encoder that intercepts such byte objects and safely decodes them into UTF-8 strings, replacing invalid characters rather than raising exceptions.
As will be clearer in the following, since our methodology leverages the EMBER feature extraction pipeline, we serialize the output of this step into the JSON format.

\subsubsection{PE Feature Extraction.}

Processing raw byte sequences directly is way too computationally heavy; we used the Endgame
Malware BEnchmark for Research (EMBER) feature extraction pipeline instead. EMBER\footnote{\url{https://github.com/elastic/ember}} is a labeled benchmark dataset
for training machine learning models to statically detect
malicious Windows portable executable files \cite{anderson2018ember}. 
Feature extraction is performed using the \texttt{LIEF} (Library to Instrument Executable Formats) framework, which enables structured parsing of PE binaries without execution. This approach allows the extraction of rich structural representations while preserving safety and scalability. In particular, each executable is mapped into a fixed-dimensional feature space of size $d = 2,381$. Formally, we define a feature extraction function
$\phi: \mathcal{X} \rightarrow \mathbb{R}^{2,381}$ which maps each executable $x \in \mathcal{X}$ to a structural feature vector $\phi(x)$.
The extracted representation includes multiple categories of static attributes, such as byte histograms, string-based statistics, section-level metadata, header information, and imported library dependencies. Overall, this yields a compact structural embedding that captures both syntactic and semantic properties of the binary.

All extracted features are serialized into JSON format for intermediate storage. However, due to the high dimensionality and scale of the dataset, loading the entire dataset into memory is not feasible. To address this limitation, we transform the dataset into memory-mapped arrays in \texttt{.dat} format, enabling direct disk access without full RAM loading. Let $N$ denote the number of samples. The dataset is represented as a matrix: $X \in \mathbb{R}^{N \times 2,381}$, which is stored in a memory-mapped format, allowing efficient random access to individual samples $x_i$ without requiring full materialization in memory.

\subsubsection{Features Preparation.}
Malware feature representations exhibit strong statistical and structural dependencies, reflecting the internal organization of PE files. Consequently, independent perturbation of individual features may disrupt these dependencies and degrade the semantic coherence of the representation space. To address this issue, we partition the extracted feature vector of dimensionality $d = 2,381$ into a set of disjoint feature groups $\mathcal{G} = \{g_1, g_2, \dots, g_K\}$, where each group $g_k \subset \mathbb{R}^{2{,}381}$ contains approximately 50 correlated features and satisfies:
\[
\bigcup_{k=1}^{K} g_k = \{1, \dots, 2{,}381\}, \quad g_i \cap g_j = \emptyset \ \text{for } i \neq j.
\]

This grouping strategy ensures that subsequent transformations operate at the granularity of semantically coherent feature blocks rather than individual features, thereby preserving structural dependencies within the representation space. A stratified train-test split is then applied to the dataset, yielding training and test sets with an 80/20 ratio while preserving class priors:
\[
\mathcal{D}_{train}, \mathcal{D}_{test} \sim P(X, Y), \quad P_{train}(Y) \approx P_{test}(Y).
\]

After these steps, we train a Base Classifier (BC) using LightGBM (Light Gradient Boosting Machine)\footnote{\url{https://lightgbm.readthedocs.io}}, a gradient-boosted decision tree framework well-suited for high-dimensional tabular data. The BC is trained on the clean training split $\mathcal{D}_{train}$ using $100$ estimators. Formally, the classifier learns a decision function $f_{BC}: \mathbb{R}^{2{,}381} \rightarrow \{0,1\}$, which approximates the optimal decision boundary under the empirical data distribution. The role of the BC is to serve as a reference baseline model that captures standard discriminative structure in the feature space. It is subsequently used to identify malicious samples for targeted augmentation in the training of the Smoothed Classifier.

\begin{figure}[ht]
    \centering
    \includegraphics[width=0.6\textwidth]{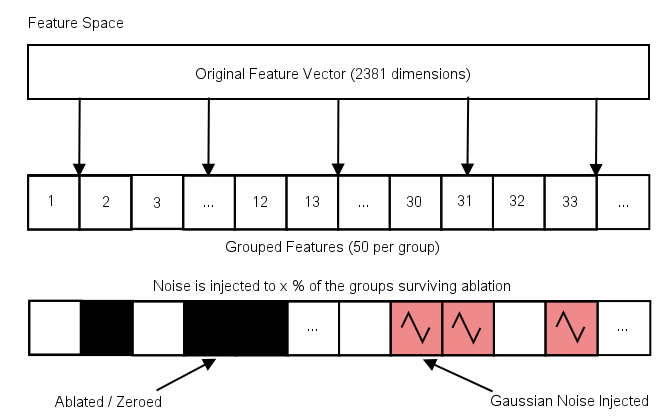}
    \caption{The Feature Mutation Process, demonstrating group-wise feature ablation and targeted noise injection for Smoothed Classifier training.}
    \label{fig:feature_mutation}
\end{figure}

To construct a robustness-enhanced model, we apply a targeted augmentation strategy focused exclusively on samples predicted as malicious by the Base Classifier. Let
$\mathcal{D}_m^{BC} = \{ x \in \mathcal{D}_{train} \mid f_{BC}(x) = 1 \}$, only samples in $\mathcal{D}_m^{BC}$ are selected for augmentation, while benign samples remain unchanged to preserve the integrity of the benign distribution. For each selected malicious sample $x$, we generate $M = 15$ stochastic variants using a two-stage transformation process as shown in Figure \ref{fig:feature_mutation}:

\begin{enumerate}
    \item \textbf{Feature Ablation.} A subset of feature groups is randomly retained. Specifically, $80\%$ of the groups are preserved, while the remaining $20\%$ are fully removed (set to zero). This operation enforces robustness to partial feature observability.

    \item \textbf{Noise Injection.} Within the retained feature groups, a subset of $10\%$ of features is selected and perturbed by additive Gaussian noise: $\epsilon \sim \mathcal{N}(0, \rho^2), \quad \rho = 0.3$. Importantly, noise is applied only to non-ablated features to avoid introducing structurally inconsistent artifacts.
\end{enumerate}

Let $\tilde{\mathcal{D}}_m$ denote the resulting augmented malicious set. The final training dataset for the Smoothed Classifier is defined as $\mathcal{D}_{SC} = \mathcal{D}_b \cup \mathcal{D}_m \cup \tilde{\mathcal{D}}_m$.

\subsubsection{Smoothed Classifier Training.} The prepared feature representations are used to train the Smoothed Classifier (SC), which is instantiated using the same LightGBM architecture as the Base Classifier (BC), but trained on an augmented dataset designed to enforce robustness. Let $\mathcal{D}_{SC}$ denote the augmented training dataset defined in the previous section. Each original sample $x \in \mathcal{D}_{SC}$ is associated with a set of perturbed instances generated through stochastic transformations combining feature-group ablation and targeted noise injection. Formally, for each input $x$, we define a transformation distribution:
\[
T(x) = \{ \tilde{x} \mid \tilde{x} = \mathcal{A}(x) + \epsilon \},
\]
where $\mathcal{A}(\cdot)$ denotes the group-wise ablation operator and $\epsilon \sim \mathcal{N}(0, \rho^2)$ represents additive Gaussian noise applied to a subset of non-ablated features.

The SC is trained to learn a decision function $f_{SC}: \mathbb{R}^{2{,}381} \rightarrow \{0,1\}$, by minimizing the empirical risk over the distribution of perturbed samples:
\[
\mathbb{E}_{(x,y) \sim \mathcal{D}_{SC}} \; \mathbb{E}_{\tilde{x} \sim T(x)} \left[ \ell\left(f_{SC}(\tilde{x}), y\right) \right],
\]
where $\ell(\cdot)$ denotes the classification loss function. This training strategy enforces invariance of the learned decision boundary under structured feature perturbations, encouraging the model to rely on stable and redundant feature patterns rather than brittle or instance-specific signals.

\subsection{Inference Phase}

At inference time, the SC evaluates each input sample through a stochastic aggregation process rather than a single deterministic forward pass. Given an input feature vector $x \in \mathbb{R}^{2{,}381}$, we define a perturbation distribution $T(x)$ consistent with the transformations applied during training, combining group-wise feature ablation and targeted Gaussian noise injection. A set of $N$ independent perturbed instances is then generated: $\tilde{x}_1, \tilde{x}_2, \dots, \tilde{x}_N \sim T(x)$, where $N = 50$ in our experimental setting. Each perturbed instance $\tilde{x}_i$ is evaluated by the trained LightGBM-based Smoothed Classifier, producing a prediction:
\[
y_i = f_{SC}(\tilde{x}_i), \quad y_i \in \{0,1\}.
\]

The final prediction is obtained through a majority voting mechanism over the $N$ stochastic evaluations:
\[
\hat{y} = \arg\max_{c \in \{0,1\}} \sum_{i=1}^{N} \mathbb{I}\left[y_i = c\right],
\]
where $\mathbb{I}(\cdot)$ denotes the indicator function.

This aggregation procedure approximates the expected classifier output under the perturbation distribution $T(x)$, yielding a smoothed decision function:
\[
g(x) = \arg\max_{c \in \{0,1\}} \mathbb{P}_{\tilde{x} \sim T(x)} \left( f_{SC}(\tilde{x}) = c \right).
\]

The stochastic voting mechanism further enables formal robustness certification through the Wilson Score Interval. In particular, this formulation provides a conservative lower bound $p_{\text{lower}}$ on the probability that the predicted majority class corresponds to the true label, even under adversarial feature perturbations.

Let $\hat{p}$ denote the empirical probability of the majority class, i.e., the fraction of perturbed samples assigned to the most frequent class:
\[
\hat{p} = \frac{1}{N} \sum_{i=1}^{N} \mathbb{I}\left[y_i = \hat{y}\right].
\]

The lower bound $p_{\text{lower}}$ is computed using the Wilson Score Interval:
\begin{equation}
    p_{\text{lower}} = \frac{\hat{p} + \frac{z^2}{2N} - z \sqrt{\frac{\hat{p}(1-\hat{p})}{N} + \frac{z^2}{4N^2}}}{1 + \frac{z^2}{N}}
\end{equation}

\noindent
where $z$ is the critical value associated with the desired confidence level (in our case, $\alpha = 0.001$, corresponding to $99.9\%$ confidence).

If $p_{\text{lower}} > 0.5$, the majority prediction is statistically guaranteed with high confidence. Based on this bound, we compute a certified robustness radius $R$:
\begin{equation}
    R = \sigma \cdot \Phi^{-1}(p_{\text{lower}})
\end{equation}

\noindent
where $\sigma$ is the noise level used during perturbation, and $\Phi^{-1}$ denotes the inverse cumulative distribution function of the standard normal distribution.

\begin{figure}[htbp]
    \centering
    \includegraphics[width=0.8\textwidth]{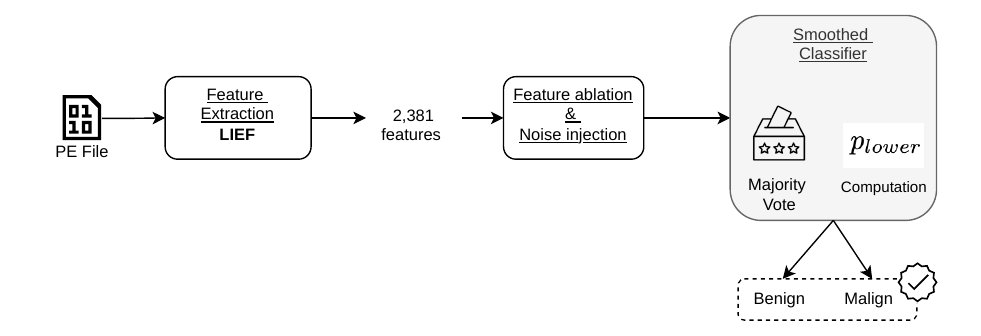}
    \caption{The Inference phase pipeline}
    \label{fig:inference_voting}
\end{figure}

To estimate the maximum robustness guarantee, we perform a systematic search over the noise parameter $\sigma$, identifying the largest value such that $p_{\text{lower}} > 0.5$ holds. This procedure yields the maximum certified robustness radius, representing the highest perturbation level under which the classifier's prediction remains provably stable.
By integrating stochastic perturbations and statistical certification, the SC produces predictions that are not only robust in practice but also accompanied by formal guarantees on their stability under adversarial feature-space transformations. Figure~\ref{fig:inference_voting} illustrates the overall inference pipeline.

\textbf{Illustrative Example}. To provide additional intuition on the certification process, we present an illustrative example.
Consider an input sample $x$ and a smoothed classifier evaluated using $N = 100$ independent perturbations (for illustrative purposes) with noise level $\sigma = 0.3$. Let $k = 90$ denote the number of perturbed instances classified as malicious. The empirical class probability is therefore $\hat{p} = \frac{k}{N} = 0.9$. To account for finite-sample uncertainty, we compute a lower confidence bound $p_{\mathrm{lower}}$ using the Wilson score interval at confidence level $1 - \alpha = 0.999$. Substituting $k = 90$, $n = 100$, and the corresponding critical value $z$, we obtain a conservative estimate $p_{\mathrm{lower}} \approx 0.78$.

Since $p_{\mathrm{lower}} > 0.5$, the prediction of the majority class is certifiably robust. The corresponding certified radius is then given by:
\begin{equation}
    R = \sigma \cdot \Phi^{-1}(p_{\mathrm{lower}}) = 0.3 \cdot \Phi^{-1}(0.78) \approx 0.23.
\end{equation}

Formally, this implies that for any perturbed input $x'$ such that $\|x' - x\|_2 < R$, the prediction of the smoothed classifier remains invariant. Hence, the classifier is provably robust within an $L_2$ ball of radius $R$ centered at $x$.

\section{Experimental Evaluation}\label{sec:result}
In this section, we evaluate the proposed framework with the goal of addressing the research questions introduced in the Introduction. All experiments were conducted in a controlled Python environment to ensure reproducibility across data preprocessing, feature extraction, model training, and evaluation.

To define the threat model, we generated adversarial data by applying PyMetaEngine to raw executables from the test split, producing functional metamorphic variants. These samples were then processed through our pipeline to obtain adversarial feature representations. In addition, to assess the generality of the proposed defense, we replicated the methodology using the MalConv architecture \cite{raff2017malware}, enabling evaluation across both traditional machine learning and deep learning models.

Given the clean and adversarial datasets, the evaluation proceeds as follows:
\begin{itemize}
\item \textbf{Baseline and Synthetic Noise Evaluation.} Performance is first assessed on clean data and under synthetic Gaussian noise to establish baseline accuracy and stability.
\item \textbf{Metamorphic Evasion Evaluation.} The classifiers are then tested against PyMetaEngine-generated samples to quantify robustness against realistic metamorphic attacks.
\item \textbf{Generalization Across Model Architectures using MalConv.} Finally, the approach is validated on an additional deep learning architecture to consolidate the obtained results.
\end{itemize}

\subsection{Baseline and Synthetic Noise Evaluation}

To establish a reliable baseline, we first evaluate the standard Base Classifier (BC) on the unmodified $20\%$ test split. The BC is implemented as a LightGBM model trained exclusively on clean data, without any form of feature ablation or noise injection. Subsequently, we assess the Smoothed Classifier (SC) on the same clean test set. During training, the SC is exposed to augmented data obtained by generating multiple perturbed variants of each malware sample. In particular, we create five clones per sample, applying random ablation to $10\%$ of the feature groups and injecting moderate Gaussian noise ($\sigma = 0.15$). This setup allows us to quantify any potential degradation in performance caused by training on corrupted inputs. At inference time, each test sample is evaluated through $N=50$ stochastic forward passes, and the final prediction is determined via majority voting. Finally, both models are evaluated under a synthetic attack scenario, where feature-space perturbations simulate adversarial conditions. The SC leverages its stochastic inference mechanism to reconstruct predictions through randomized smoothing.

As reported in Table~\ref{tab:performanceComp}, the BC achieves near-perfect performance on the clean dataset, with an accuracy of 99.75\% and a recall of 99.77\%. This confirms both the high quality of the dataset and the effectiveness of structural EMBER features for malware detection under standard conditions. The corresponding confusion matrix (see Figure~\ref{fig:confusion_matrices_all} in the Appendix) further highlights the negligible number of misclassifications. When evaluated on the same clean data, the SC does not exhibit any performance degradation. On the contrary, it slightly improves over the baseline, reaching an accuracy of $99.81\%$. This indicates that the proposed group-wise ablation strategy preserves the structural semantics of the feature space while enhancing generalization. Under synthetic perturbations, the BC shows a marked degradation in performance. As shown in Table~\ref{tab:performanceComp}, the accuracy drops to $96.28\%$, while recall decreases to $95.73\%$. The reduction in recall is particularly critical, as it corresponds to an increase in false negatives, meaning that malicious samples are more likely to be misclassified as benign. In contrast, the SC demonstrates strong resilience to feature-space perturbations. Despite the adversarial setting, it maintains an accuracy of $99.53\%$ and a recall of $99.75\%$, effectively preserving its detection capability. This robustness can be attributed to the training procedure, which enforces invariance to partial and noisy feature representations. As a result, the synthetic perturbations fail to shift samples across the smoothed decision boundary. The corresponding confusion matrix (Figure~\ref{fig:confusion_matrices_all} in Appendix) confirms this behavior, showing performance levels comparable to those observed on clean data.

\begin{table}
    \centering
    \scriptsize
    \caption{Performance comparison of Base Classifier (BC) and Smoothed Classifier (SC) under 0\% and 10\% noise conditions.}
    \label{tab:performanceComp}
    \begin{tabular}{llcccc}
        \toprule
        \textbf{Noise} & \textbf{Scenario} & \textbf{Accuracy} & \textbf{Precision} & \textbf{Recall} & \textbf{F1-Score} \\
        \midrule
        \multirow{2}{*}{0\%} 
        & BC & 99.75\% & 99.72\% & 99.77\% & 99.74\% \\
        & SC & 99.81\% & 99.75\% & 99.87\% & 99.81\% \\
        \midrule
        \multirow{2}{*}{10\%} 
        & BC & 96.28\% & 96.70\% & 95.73\% & 96.21\% \\
        & SC & 99.53\% & 99.30\% & 99.75\% & 99.53\% \\
        \bottomrule
    \end{tabular}
\end{table}

To thoroughly understand the limits of our models, we decided to push the synthetic evaluation into a multi-level stress test. Rather than stopping at a single noise threshold, we wanted to see exactly how much structural damage it would take to break the classifiers. We systematically escalated the attack by adding noise to 20\%, 30\%, and finally 40\% of the feature groups. 

The results of this stress test, detailed in Table~\ref{tab:multilevel_noise}, are highly revealing. As the noise level scales to 40\%, the unprotected Base Classifier suffers a catastrophic failure. Its recall plummets to 86.97\%, meaning the model becomes functionally blind to over 13\% of the malware samples, letting them bypass the system. In stark contrast, the Smoothed Classifier absorbs the massive feature-space damage with ease. Even with 40\% of the executable's structural data corrupted or missing, the SC maintains a near-perfect 99.62\% recall and 99.32\% accuracy. This experiment definitively proves that our feature-group voting mechanism does not just survive minor statistical drift, but provides an incredibly deep buffer against severe adversarial degradation.

\textbf{Findings.} All the above results support \textbf{RQ2}, showing that randomized smoothing improves robustness without sacrificing accuracy: the SC remains stable under perturbations, while the BC degrades. Additionally, the consistent behavior of the SC across noisy evaluations provides evidence for \textbf{RQ3}, supporting statistical robustness certification via majority voting and the Wilson score interval.

\begin{table}
    \centering
    \scriptsize
    \caption{Performance degradation under a multi-level synthetic noise stress test. {\em Base Classifier (BC); Smoothed Classifier (SC)}}
    \label{tab:multilevel_noise}
    \begin{tabular}{llcccc}
        \toprule
        \textbf{Noise} & \textbf{Scenario} & \textbf{Accuracy} & \textbf{Precision} & \textbf{Recall} & \textbf{F1-Score} \\
        \midrule
        \multirow{2}{*}{20\%} 
        & BC & 95.72\% & 96.36\% & 94.93\% & 95.64\% \\
        & SC & 99.57\% & 99.36\% & 99.77\% & 99.56\% \\
        \midrule
        \multirow{2}{*}{30\%} 
        & BC & 93.26\% & 94.67\% & 91.51\% & 93.06\% \\
        & SC & 99.38\% & 99.09\% & 99.66\% & 99.37\% \\
        \midrule
        \multirow{2}{*}{40\%} 
        & BC & 90.28\% & 92.90\% & 86.97\% & 89.84\% \\
        & SC & 99.32\% & 99.02\% & 99.62\% & 99.32\% \\
        \bottomrule
    \end{tabular}
\end{table}

\subsection{Metamorphic Evasion Evaluation}

To evaluate robustness against realistic adversarial threats, we conducted experiments using PyMetaEngine on malware-only test samples, focusing on evasion (i.e., false negatives). Since only malicious instances are considered, recall is the primary metric. Mutated binaries were generated and processed through the EMBER pipeline to extract adversarial feature vectors. Without additional perturbations, both the Base Classifier (BC) and Smoothed Classifier (SC) maintain high recall ($99.80\%$ and $99.86\%$, respectively), indicating that metamorphic transformations alone are insufficient to significantly alter high-level structural features. We then introduced a stronger combined attack by adding synthetic noise (10\% feature-group ablation with Gaussian noise, $\sigma = 0.3$). Under this setting, the BC shows a clear degradation (recall drops to $94.77\%$), whereas the SC remains unaffected, maintaining a recall of $99.86\%$, thus demonstrating strong resilience to combined adversarial perturbations.
To further assess robustness, we conduct a multi-level stress test by increasing noise to 20\%, 30\%, and 40\% of feature groups (Table~\ref{tab:pymeta_stress}). The BC degrades progressively, reaching $86.97\%$ recall at 40\% noise, while the SC remains highly stable ($99.62\%$ recall), even under severe perturbations. This highlights the ability of the smoothing mechanism to preserve decision consistency despite substantial structural degradation.

\begin{table}
    \centering
    \scriptsize
    \caption{Performance comparison on mutated data: Pure PyMetaEngine (0\% Noise) vs. Combined Attack (10\% Noise).{\em Base Classifier (BC); Smoothed Classifier (SC)}}
    \label{tab:pymeta_initial}
    \begin{tabular}{llcccc}
        \toprule
        \textbf{Noise} & \textbf{Scenario} & \textbf{Accuracy} & \textbf{Precision} & \textbf{Recall} & \textbf{F1-Score} \\
        \midrule
        \multirow{2}{*}{0\%} 
        & BC & 99.80\% & 100.00\% & 99.80\%  & 99.90\% \\
        & SC & 99.86\% & 100.00\% & 99.86\%  & 99.93\% \\
        \midrule
        \multirow{2}{*}{10\%} 
        & BC & 94.77\% & 100.00\% & 94.77\%  & 97.31\% \\
        & SC & 99.86\% & 100.00\% & 99.86\%  & 99.93\% \\
        \bottomrule
    \end{tabular}
\end{table}

\begin{table}
    \centering
    \scriptsize
    \caption{Stress test performance under escalated combined attacks (20\%, 30\%, and 40\% Noise). {\em Base Classifier (BC); Smoothed Classifier (SC)}}
    \label{tab:pymeta_stress}
    \begin{tabular}{llcccc}
        \toprule
        \textbf{Noise} & \textbf{Scenario} & \textbf{Accuracy} & \textbf{Precision} & \textbf{Recall} & \textbf{F1-Score} \\
        \midrule
        \multirow{2}{*}{20\%} 
        & BC & 91.92\% & 100.00\% & 91.92\%  & 95.79\% \\
        & SC & 99.78\% & 100.00\% & 99.78\%  & 99.89\% \\
        \midrule
        \multirow{2}{*}{30\%} 
        & BC & 87.92\% & 100.00\% & 87.92\%  & 93.57\% \\
        & SC & 99.70\% & 100.00\% & 99.70\%  & 99.85\% \\
        \midrule
        \multirow{2}{*}{40\%} 
        & BC & 84.14\% & 100.00\% & 84.14\%  & 91.39\% \\
        & SC & 99.60\% & 100.00\% & 99.60\%  & 99.80\% \\
        \bottomrule
    \end{tabular}
\end{table}

\textbf{Findings.} The above results demonstrate that metamorphic transformations significantly degrade the performance of the BC, confirming the vulnerability of standard feature-based approaches to evasion techniques (supporting \textbf{RQ1}). In contrast, the SC demonstrates strong resilience, with substantially lower performance degradation. This confirms that feature-based randomized smoothing improves empirical robustness against realistic metamorphic attacks, providing clear evidence in support of \textbf{RQ2}.

\subsection{Generalization Across Model Architectures using MalConv}

This experiment evaluates whether the proposed randomized smoothing defense generalizes beyond tree-based models to fundamentally different architectures. To this end, we adapt our approach to MalConv, a deep convolutional neural network that operates directly on raw byte sequences, bypassing manual feature extraction. Executable files are processed as byte streams truncated to 2MB, where each byte is represented as an integer in $[0,255]$ and padded with a special token when necessary. Since noise cannot be directly applied to discrete inputs, we inject Gaussian noise ($\sigma=0.3$) in the continuous embedding space, where bytes are mapped to 8-dimensional vectors. This strategy enforces robustness by preventing the model from relying on brittle byte-level patterns.

We train both a baseline Base Classifier (BC) and a Smoothed Classifier (SC) using this noisy embedding mechanism. As shown in Table~\ref{tab:malconv_results}, the SC preserves performance on clean data ($98.33\%$ accuracy vs. $98.70\%$ for BC), indicating that robustness does not come at the cost of accuracy. Under a $10\%$ byte corruption attack, the BC exhibits degraded precision ($94.88\%$), reflecting increased false positives. In contrast, the SC maintains stable performance ($98.56\%$ accuracy), demonstrating effective mitigation of perturbations through stochastic voting.

This robustness is further supported by the confusion matrices in Figure~\ref{fig:malconv_cm} and Venn diagrams in Figure~\ref{fig:malconv_venn} in the Appendix, which show that the SC significantly reduces noise-induced misclassifications while maintaining a consistent error profile. After this experiment, to stress-test both models, we increase corruption levels to 20\%, 30\%, and 40\%. As reported in Table~\ref{tab:malconv_multilevel_noise}, the BC progressively collapses, reaching $65.54\%$ accuracy at 40\% noise and exhibiting a degenerate behavior (near-total malware predictions). Conversely, the SC maintains strong performance even under extreme perturbations ($94.76\%$ accuracy), confirming the effectiveness of the smoothing mechanism.

\textbf{Findings.} These results address \textbf{RQ4} by demonstrating that feature-space randomized smoothing is not restricted to tabular data or decision-tree models. It operates as a highly adaptable structural defense, providing empirical robustness to deep neural networks evaluating raw binary sequences.

\begin{table}
    \centering
    \scriptsize
    \caption{Performance of MalConv architectures on clean data and under a 10\% raw byte corruption attack.}
    \label{tab:malconv_results}
    \begin{tabular}{llcccc}
        \toprule
        \textbf{Noise} & \textbf{Scenario} & \textbf{Accuracy} & \textbf{Precision} & \textbf{Recall} & \textbf{F1-Score} \\
        \midrule
        \multirow{2}{*}{0\%} 
        & Base MalConv     & 98.70\% & 98.12\% & 99.26\% & 98.69\% \\
        & Smoothed MalConv & 98.33\% & 99.49\% & 97.11\% & 98.29\% \\
        \midrule
        \multirow{2}{*}{10\%} 
        & Base MalConv     & 97.18\% & 94.88\% & 99.66\% & 97.21\% \\
        & Smoothed MalConv & 98.56\% & 99.23\% & 97.83\% & 98.53\% \\
        \bottomrule
    \end{tabular}
\end{table}

\begin{table}
    \centering
    \scriptsize
    \caption{Performance degradation of MalConv architectures under a multi-level raw byte corruption stress test (20\%, 30\%, 40\%).}
    \label{tab:malconv_multilevel_noise}
    \begin{tabular}{llcccc}
        \toprule
        \textbf{Noise} & \textbf{Scenario} & \textbf{Accuracy} & \textbf{Precision} & \textbf{Recall} & \textbf{F1-Score} \\
        \midrule
        \multirow{2}{*}{20\%} 
        & Base MalConv     & 91.93\% & 86.06\% & 99.83\%  & 92.44\% \\
        & Smoothed MalConv & 98.59\% & 98.54\% & 98.61\%  & 98.58\% \\
        \midrule
        \multirow{2}{*}{30\%} 
        & Base MalConv     & 80.64\% & 71.85\% & 99.96\%  & 83.60\% \\
        & Smoothed MalConv & 98.06\% & 96.94\% & 99.20\%  & 98.06\% \\
        \midrule
        \multirow{2}{*}{40\%} 
        & Base MalConv     & 65.54\% & 58.90\% & 100.00\% & 74.14\% \\
        & Smoothed MalConv & 94.76\% & 90.72\% & 99.56\%  & 94.94\% \\
        \bottomrule
    \end{tabular}
\end{table}

\section{Conclusion}
\label{sec:conclusion}

Malware detection systems based on machine learning are increasingly challenged by adversarially modified executables and structurally preserved evasion techniques. In this work, we proposed a robust malware classification framework based on randomized feature smoothing, where grouped feature perturbations are introduced during both training and inference to improve stability in high-dimensional feature spaces. Our approach combines stochastic voting with statistical certification via the Wilson Score Interval, enabling probabilistic guarantees on prediction reliability under bounded perturbations. Experimental results demonstrate that the proposed Smoothed Classifier maintains strong performance on clean data while significantly improving robustness under both synthetic feature-space noise and realistic metamorphic transformations. In particular, the model effectively mitigates the degradation observed in standard classifiers when exposed to structured perturbations. Furthermore, the certification framework provides meaningful lower bounds on prediction confidence and allows for the estimation of input-specific robustness radii through a binary search procedure over the noise scale. The evaluation across both LightGBM and MalConv architectures further confirms that the proposed methodology is architecture-agnostic and can be integrated into both traditional machine learning pipelines and deep learning-based malware detectors without compromising baseline accuracy. 

Despite these promising results, several directions remain open for future work. For instance, the current smoothing strategy relies on predefined feature grouping, which could be improved through adaptive or learned grouping mechanisms that better capture semantic dependencies in executable features. Additionally, while our evaluation includes synthetic and metamorphic transformations, future work should investigate more complex real-world adversarial malware families and evolving attack strategies.

\bibliographystyle{plain}
{\footnotesize
\bibliography{biblio}
}

\appendix
\section{Appendix}

This appendix reports additional experiment results.

\subsection{Baseline and Synthetic Noise Evaluation}
The confusion matrices show that both models achieve near-perfect classification on clean data, with strongly diagonal patterns. Under synthetic perturbations, the Base Classifier degrades, exhibiting more false negatives, while the Smoothed Classifier maintains a stable and well-separated decision boundary across conditions.

\begin{figure}
    \centering
    \includegraphics[width=1\textwidth]{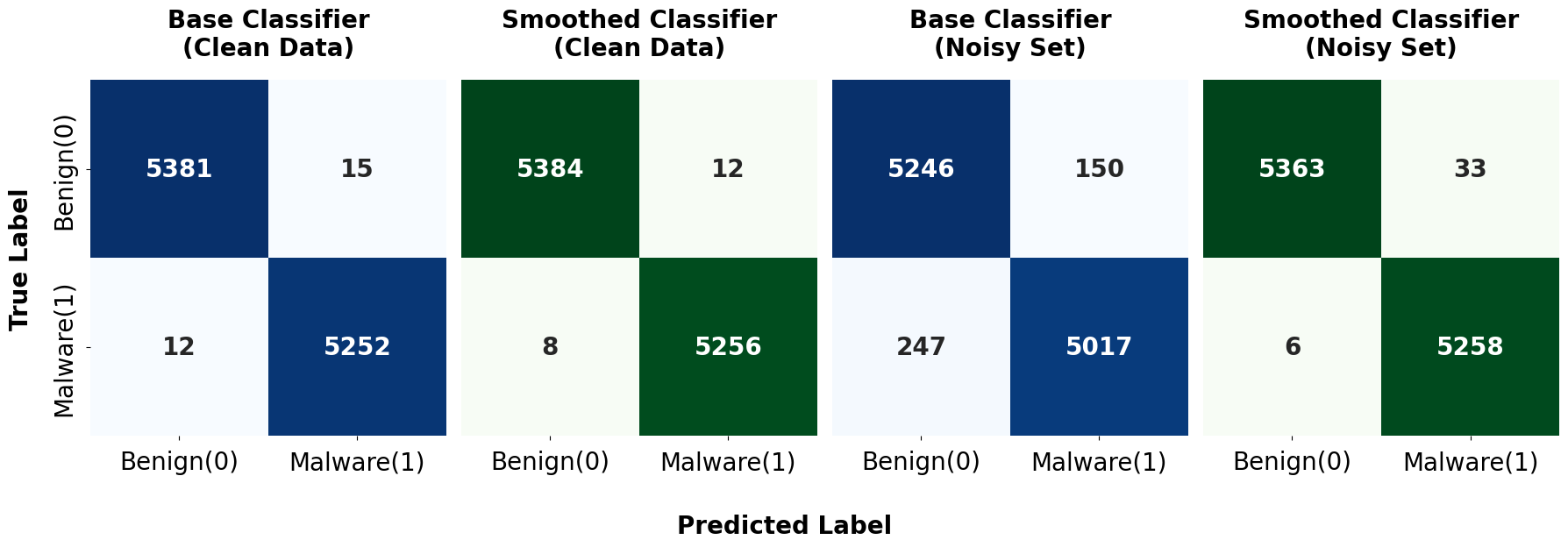}
    \caption{Confusion matrices comparing Base Classifier (BC) and Smoothed Classifier (SC) under clean and synthetic noise conditions.}
    \label{fig:confusion_matrices_all}
\end{figure}

\subsection{Metamorphic Evasion Evaluation}
In the following experiment, we confirm that PyMetaEngine produces genuine binary modifications by comparing SHA-256 hashes and analyzing the resulting EMBER feature vectors. Despite these changes, cosine similarity reveals that most transformations are micro-mutations, preserving over 99\% structural similarity. 

\begin{figure}
    \centering
    \includegraphics[width=0.8\textwidth]{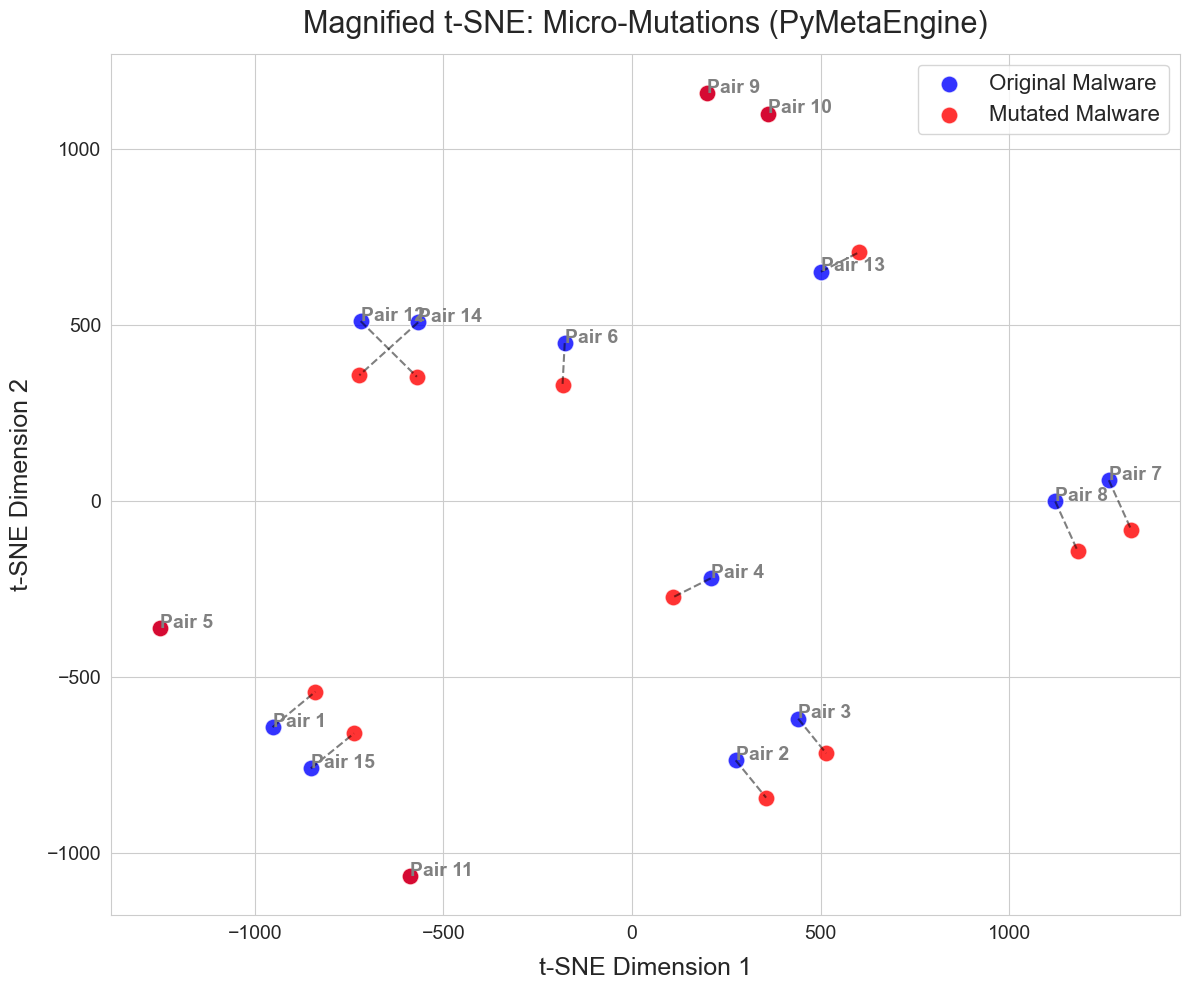}
    \caption{Magnified t-SNE projection of 15 micro-mutated pairs, illustrating the exact perturbation trajectory between the original executable and its metamorphic clone.}
    \label{fig:tsne_micro}
\end{figure}
To inspect the exact nature of this evasion trajectory, we isolated 15 specific pairs of these micro-mutations and ran a highly localized t-SNE projection (configured with a low perplexity score to magnify local distances). Figure~\ref{fig:tsne_micro} illustrates this magnified view, drawing a direct vector line between the original file and its metamorphic clone. This mapping visualizes the exact adversarial distance that the PyMetaEngine generates, representing the precise mathematical gap our Smoothed Classifier must learn to bridge through its voting mechanism.
Whereas the empirical analysis has shown the resistance of the EMBER-trained Smoothed Classifier to high-amplitude attacks on the structure of the malware, the major strength of the proposed solution is in its potential for mathematical proof. The mathematical proof is possible through analysis of the voting pattern formed during the stochastic phase and calculation of a certified radius $R$ within the EMBER feature space. For each sample individually, this radius defines the bounded region within which the classification of the malware is mathematically guaranteed to remain stable, no matter what metamorphic transformations the PyMetaEngine attack performs.

\begin{figure}
    \centering
    \includegraphics[width=0.85\textwidth]{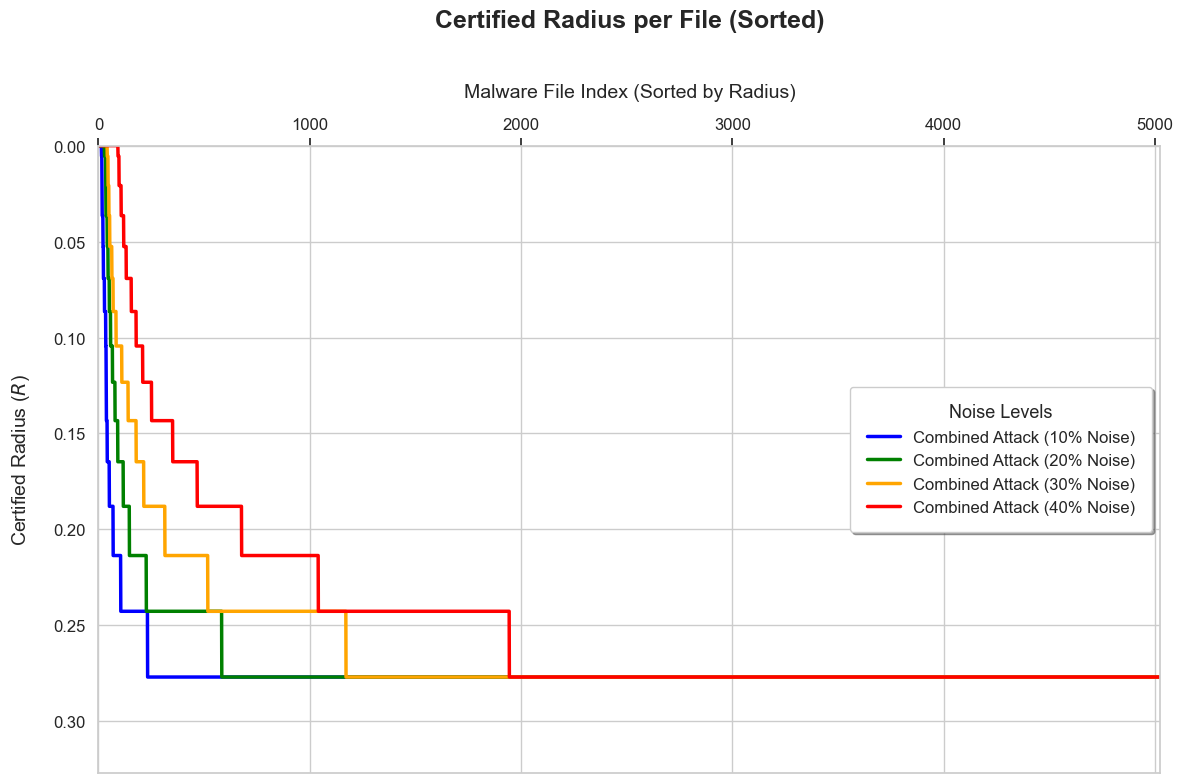}
    \caption{Demonstrating the certified radius, calculated by the EMBER-trained Smoothed Classifier, across various PyMetaEngine mutated samples.}
    \label{fig:certified_curve}
\end{figure}

\subsection{Generalization Across Model Architectures using MalConv}

\begin{figure}
    \centering
    \includegraphics[width=1\textwidth]{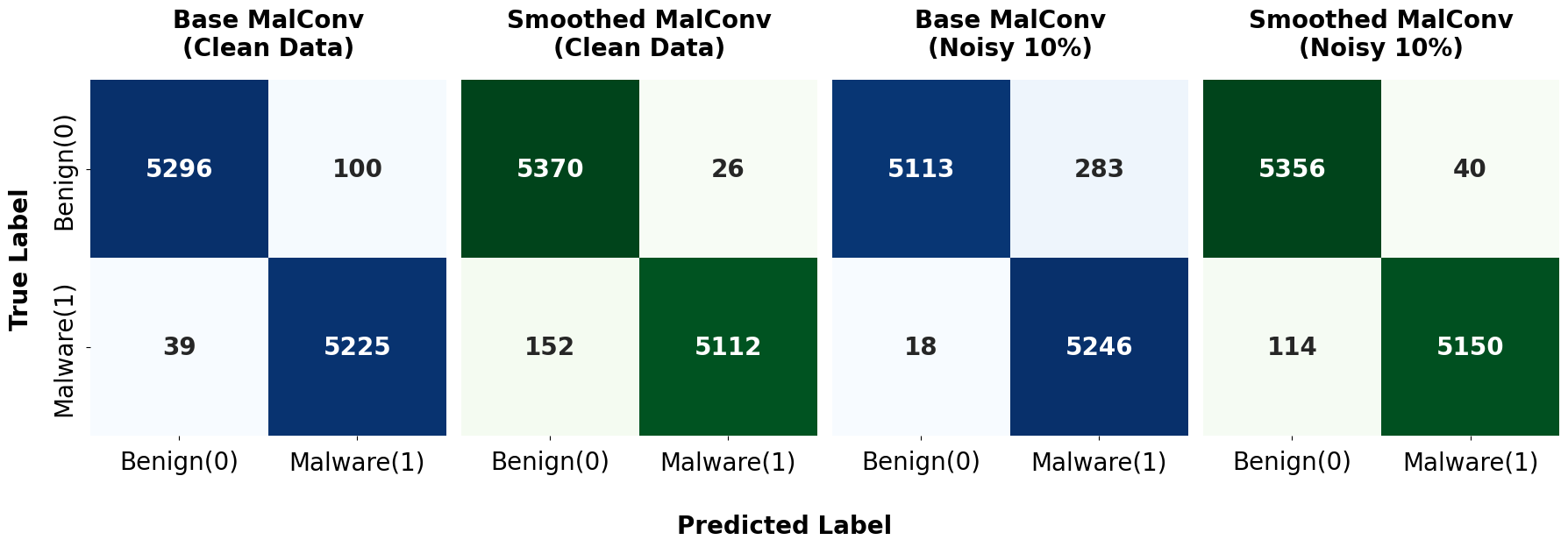}
    \caption{Confusion Matrices comparing the Base and Smoothed MalConv architectures on clean data and under the 10\% byte corruption attack.}
    \label{fig:malconv_cm}
\end{figure}

\begin{figure}
    \centering
    \includegraphics[width=0.9\textwidth]{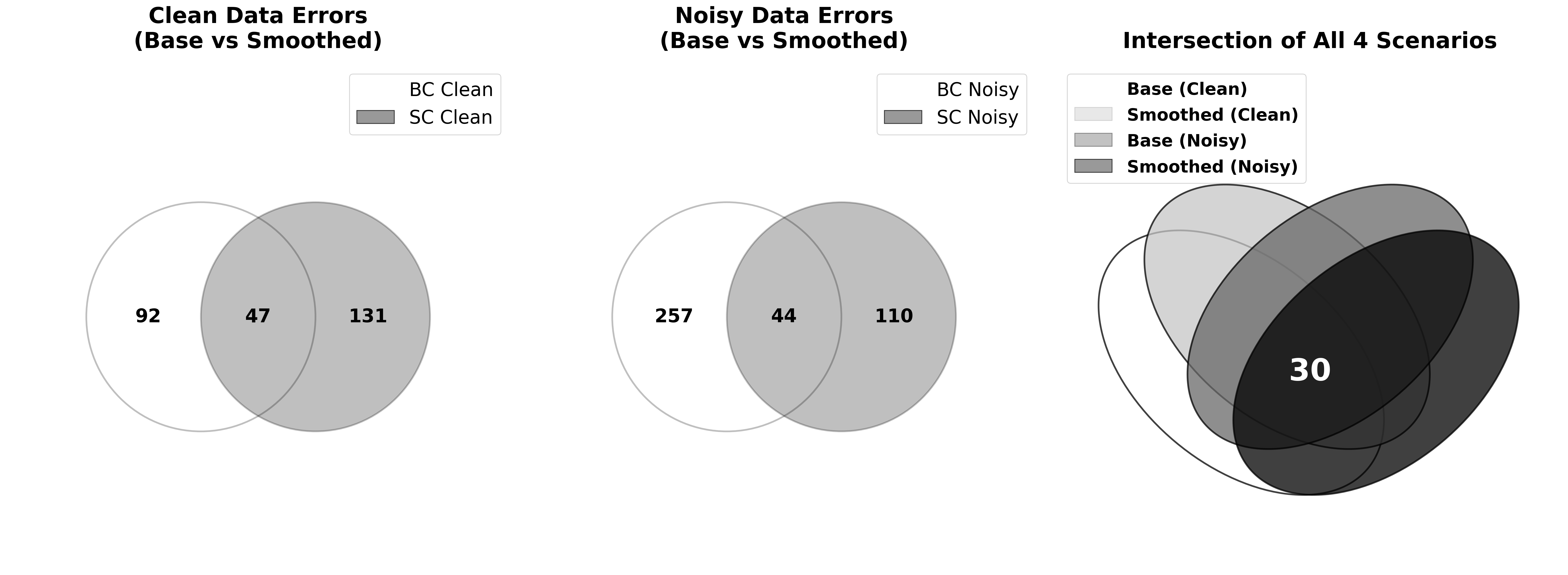}
    \caption{Intersection of misclassified samples across the baseline and smoothed MalConv architectures, highlighting error distribution shifts under adversarial noise.}
    \label{fig:malconv_venn}
\end{figure}

The confusion matrices reported in Figure~\ref{fig:malconv_cm} provide a detailed visualization of the classification behavior of the MalConv Base Classifier (BC) and Smoothed Classifier (SC) under both clean and adversarial conditions. Under clean data, both models exhibit a strongly diagonal structure in their confusion matrices, indicating highly accurate separation between benign and malicious executables.
Under the 10\% byte corruption attack, a clear divergence emerges between the two architectures. The confusion matrix of the BC shows an increase in misclassified samples, whereas SC maintains a highly concentrated diagonal structure, with only minimal misclassifications. This reflects the effectiveness of the randomized smoothing mechanism and majority voting strategy in mitigating the impact of low-level structural noise. The Venn diagrams reported in Figure~\ref{fig:malconv_venn} illustrate the overlap of misclassified samples across clean and adversarial evaluation scenarios for both the Base Classifier (BC) and the Smoothed Classifier (SC). This representation provides additional insight into the stability and consistency of the decision boundaries under distributional shifts induced by byte-level corruption. The BC exhibits a highly volatile error distribution under attack; the SC maintains a consistent and constrained set of misclassifications, further supporting its robustness to adversarial byte-level perturbations.

\end{document}